# Rejoinder: Expert Elicitation for Reliable System Design

**Tim Bedford, John Quigley and Lesley Walls**

First of all, we would like to thank the discussants for the care and thoughtfulness that they have taken in preparing their comments.

Koehler presents a helpful discussion, putting forward a number of different ideas that generalize the approach taken. A taxonomy for technical system elicitation would provide useful guidance for practitioners and serve to codify applicable assumptions during the different systems engineering phases. Although more research is needed here, one could see the emergence of international standards that rely on such a taxonomy.

We acknowledge that the elicitation problem varies greatly depending on the technical system as pointed out by Koehler and we have sought to generalize our experience in studying complex systems, including aerospace, rail and naval for both commercial and defense markets. This explains our bias toward the "closed loop" case. We agree with the two extra areas of expert elicitation identified for "waterfall" cases: lack of expertise continuity and the problem of "forward casting" requirements for an existing system. Both of these relate to discontinuous changes in system operation. Such changes have occurred most obviously in military systems and other projects with long lead times. However, in the commercial world, such discontinuities can be forced by regulatory or market changes, or by outsourcing decisions. These may make historic data collection taxonomies less relevant to the reliability questions posed to support new operational decisions and, therefore, provide new areas of application for expert judgement techniques.

The final point raised by Koehler about the difficulties imposed by system complexity is well made and the notion of multiple concurrent reliability models is intriguing. This does partially link into the notion of expert weighting. However, it also requires a good understanding of the notion of model "expertise" as distinct from expert "expertise." One might argue that if sufficient understanding exists to be able to quantify model expertise, then one should be able to directly build a meta model that incorporates the best of each model. In practice, the need to be cost-efficient will usually mitigate against such a strategy, and model combination is an interesting alternative.

Wang rightly observes that we have not tried to give a survey of expert judgement methodologies. The main reason for this is that several surveys have been undertaken, including a recent one with a wide coverage (Jenkinson, 2005). It has not been our purpose to survey these methods again. Instead we aim to discuss the context in which such models may be used in the engineering design process and to show that the expert problem in this context frequently is more demanding than a "straightforward" probability elicitation.

Having said this, Wang is right to identify empirical Bayes (EB) as an interesting method with potential application in the area under discussion. There is, however, more than one way to utilize this approach. The approach discussed by Wang explicitly uses expert information as data, hence forcing the analyst to choose priors and likelihoods for the expert data given the parameters. This is a fundamental problem because it forces the analyst into the role of meta expert. In this case, the specification of $p(x|\Theta)$ is going to be problematic whether or not we use EB. In our own work with EB (Quigley, Bedford and Walls, 2006, 2007) we have integrated expert judgement into the approach through the selection of pools that comprise different types of events whose data are merged in the EB process. The use of EB allows us to increase the quantity of data available to make estimates of reliability parameters through expert judgements about which events should have similar order of magnitude behavior.

Wang's proposal for using evidential reasoning in reliability combines a number of different questionable features. For the purposes of this rejoinder, we propose distinguishing three different issues contained in the discussion:

- Nonprobabilistic representations of uncertainty.
- Imprecise uncertainties.
- Multicriteria decision models.

*Nonprobabilistic representations of uncertainty*: We are yet to be convinced that these play a useful role. The examples we have seen discussed—both examples to show the limitations of probability and examples to show the need for a more general framework—are marred by lack of clarity about the underlying problem being modeled. Indeed this sometimes seems to be the point of the "need" for something else. In many cases more attention paid to structuring the problem and articulating the reasons for modeling will surely take care of many of the ambiguities. To paraphrase O'Hagan and Oakley ([2004](#)), who recently wrote a paper titled "Probability is perfect, but we can't elicit it perfectly," we might say that "probability is perfect, but we find it difficult to apply appropriately." Such difficulties are even more apparent when applied to more complex generalizations of probability. The danger is that theoreticians use such methods as a fix to avoid resolving important modeling issues.

*Imprecise uncertainties*: There is growing interest, and some sound foundational work, in the area of interval probabilities. Such quantities may have a real and useful application, particularly in bounding probabilities of undesirable events. See, for example, Coolen, Coolen-Schrijner and Yan ([2002](#)), Coolen and Yan ([2003](#)), Coolen ([2004](#), [2006](#)), Augustin and Coolen ([2004](#)) and Coolen and Coolen-Schrijner ([2005](#)).

*Multicriteria decision models*: It is important not to confuse such models, which in the first instance are designed to represent trade-offs between different attributes of a decision consequence, with probabilistic models that represent system and knowledge relationships. In the case of the motorcycle mentioned in the discussion, the motorcycle is modeled most simply as a series system in the subsystems mentioned. The discussion of this example seems to force the analyst down a more complex route that ignores the basic engineering structure of the system.

Furthermore, so many elements of the calculation appear to be arbitrary—for example, what is the event "that the $i$th basic attribute supports the hypothesis that the general attribute is assessed to the $n$th grade" that is being ascribed a probability and why should weights from Saaty's analytic hierarchy process be used to multiply probabilities?—that it is difficult to see that this leads to something really meaningful and of more use than other simpler rule-of-thumb evaluations.

The experience of Fenton and Neil in developing Bayesian methods, especially Bayesian networks, adds valuable support to many issues raised in the paper. We would certainly acknowledge that TRACS is an early example of a meta modeling system of the type we discuss and it is good to hear that model building in its more recent developments is faster. Unfortunately, because these are commercial systems, it is difficult for academics to be able to make judgements about the internal workings of the systems.

We agree with the point raised by Fenton and Neil that the customer can be an expert, as well as client, because it will often be the case that the customer possesses expertise about, for example, the operational environment and maintenance of the family of systems. Hence the boundaries between the manufacturer and customer classes in Table 1 should be taken as an example of typical stakeholder roles rather than as a fixed allocation appropriate for all systems. In those cases where the customer has dual roles, additional care is required to manage bias that arises due to the levels of trust. Our limited experience to date in working with teams that span stakeholder classes has been mixed: we have experienced a lack of openness in some situations, while in others we enjoyed a sharing in both directions motivated by the need for a useful decision support tool. The presence of trust will be influenced by the culture of the companies involved as well as the expected longevity of the relationship. The awareness and management of subjective bias is important, but we agree that it should not be regarded as a reason not to conduct Bayesian modeling.

In the absence of much relevant empirical data, Fenton and Neil point out that reliability assessment can be regarded a "black art." Certainly, Bayesian modeling can help to make assumptions more transparent. However, to some extent this simply brings with it a shift of difficulty from one area of modeling to another. The parties have to find some level of agreement on prior distributions, which can be problematic if the parties really understand the significance of the choice being made.

Fenton and Neil give examples of the use of expert elicitation within six-sigma approaches. This is noteworthy given that many reliability problems arise from systematic design variation due to management as well as technical considerations. Despite the strong relationship between reliability and quality, culturally they can be disparate within organizations. By focusing on failure mode identification and tracking, we have experienced limited success in conceptually reeliciting priors for reliability modeling using production experience (Walls, Quigley and Marshall, 2006). The reasons for only limited success can be partially attributed to common process drivers identified by the aerospace companies involved in modeling. For example, the difficulties of using standard data-driven statistical process control for low-volume manufacturing has facilitated rather than hindered the acceptance of elicitation. However, we emphasize that the conceptual acceptance by stakeholders as evidence of success in use currently remains scarce. Hence the research questions posed concerning cultural conflict, organizational drivers and process drivers are important to address issues for which only piecemeal anecdotal evidence currently exists.

We would like to clarify to Fenton and Neil that we are *not* assuming implicitly or otherwise that the benefits of probability elicitation only accrue in situations where there is already a highly developed reliability methodology and we do agree that elicitation plays a distinctive role in organizations where it is not cost-effective to collect empirical data. However, in situations where a highly developed reliability culture exists, there is a critical need to structure the models being quantified, and the users will certainly benefit from that structuring phase, as well as the later quantification.

Fenton and Neil point out that the "additional key benefit" of this kind of probability elicitation in terms of providing codified information for future systems is one that is certainly of importance in those industries with very short development cycles. For systems with longer cycles, there is time to collect operational information to update or replace the expert derived data, and industry "generic data bases" play the role discussed.

We are grateful to the discussants for their comments, which provide further insights into many issues raised in the paper and contribute a number of new ideas that were not explored within the original paper.

## REFERENCES

Augustin, T. and Coolen, F. P. A. (2004). Nonparametric predictive inference and interval probability. *J. Statist. Plann. Inference* **124** 251–272. MR2080364

Coolen, F. P. A. (2004). On the use of imprecise probabilities in reliability. *Quality and Reliability Engineering International* **20** 193–202.

Coolen, F. P. A. (2006). On nonparametric predictive inference and objective Bayesianism. *J. Logic, Language and Information* **15** 21–47. MR2254566

Coolen, F. P. A. and Coolen-Schrijner, P. (2005). Nonparametric predictive reliability demonstration for failure-free periods. *IMA J. Manag. Math.* **16** 1–11. MR2124165

Coolen, F. P. A., Coolen-Schrijner, P. and Yan, K. J. (2002). Nonparametric predictive inference in reliability. *Reliability Engineering and System Safety* **78** 185–193.

Coolen, F. P. A. and Yan, K. J. (2003). Nonparametric predictive inference for grouped lifetime data. *Reliability Engineering and System Safety* **80** 243–252.

Jenkinson, D. (2005). The elicitation of probabilities—a review of the statistical literature. BEEP working paper, Univ. Sheffield.

O'Hagan, A. and Oakley, J. E. (2004). Probability is perfect, but we can't elicit it perfectly. *Reliability Engineering and System Safety* **85** 1–3, 239–248.

Quigley, J., Bedford, T. and Walls, L. (2006). Fault tree inference for one-shot devices using Bayes and empirical Bayes methods. In *Proc. ESREL 2006 Safety and Reliability Conference* 859–865.

Quigley, J., Bedford, T. and Walls, L. (2007). Estimating rate of occurrence of rare events with empirical Bayes: A railway application. *Reliability Engineering and System Safety* **92**. To appear.

Walls, L., Quigley, J. and Marshall, J. (2006). Modeling to support reliability enhancement during product development with applications in the UK aerospace industry. *IEEE Trans. Engineering Management* **53** 263–274.